\documentclass[aps,twocolumn,floats,pre]{revtex4}
\usepackage{graphics,graphicx,epsfig}
\usepackage{amssymb}
\usepackage{epsf,epstopdf,wrapfig}

\begin{document}

\title{Rediscovering the power of pairwise interactions}

\author{William Bialek$^a$ and Rama Ranganathan$^b$}

\affiliation{$^a$Joseph Henry Laboratories of Physics, Lewis--Sigler Institute for Integrative Genomics, and\\
Princeton Center for Theoretical Physics,
Princeton University,
Princeton, New Jersey 08544\\
$^b$Howard Hughes Medical Institute, Green Center for Systems Biology, and Department of Pharmacology\\
University of Texas Southwestern Medical Center, Dallas, Texas 75390}

\date{\today}

\begin{abstract}
Two recent streams of work suggest that pairwise interactions may be sufficient to capture the complexity of biological systems ranging from protein structure to networks of neurons.  In one approach, possible amino acid sequences in a family of proteins are generated by Monte Carlo annealing of a `Hamiltonian' that forces pairwise correlations among amino acid substitutions to be close to the observed correlations.  In the other approach, the observed correlations among  pairs of neurons are used to construct a maximum entropy model for the states of the network as a whole.  We show that, in certain limits, these two approaches are mathematically equivalent, and we comment on open problems suggested by this framework.
\end{abstract}

\maketitle

\section{Introduction}

In systems  composed of many elements, rich and complex behavior can emerge from simple interactions.  Indeed, for many systems studied by physicists and chemists, we can understand almost everything by thinking just about interactions between pairs of elements.  Sometimes this is an essentially exact statement:  (almost) all the complexity of chemical bonding and reactivity has its origins in the Coulomb interactions among electrons and protons, and the total energy associated with this interaction is a sum over pairs of particles.  Alternatively, the pairwise description might not be microscopically exact, but could still be a very good approximation, as in the description of many different kinds of magnets (ferromagnets, antiferromagnets, spin glasses) using only interactions between pairs of spins.   In fact, pairwise interactions can generate essentially unlimited complexity, since finding the ground state of a model magnet with arbitrary pairwise interactions among the spins is an NP--complete problem \cite{barahona_82}.

Could models based on pairwise interactions be powerful enough to capture the behavior of biological systems?   One the one hand we know that pairwise interactions can provide considerable explanatory power, but on the other hand restricting our description to pairwise interactions is an enormous simplification.  The tension between the physicists' desire for simplification and the biologists' appreciation of complexity is the subject of well known jokes \cite{horse}.
Jokes aside, biological systems often have many elements with no obvious geometrical arguments for simplification, and certainly there are many cases where the elements (bases along DNA, amino acids along a single protein chain, proteins, cells,  ...) have many opportunities to interact in combinatorial fashion as they generate biological function.   Two recent streams of work have led to a re--examination of these issues.

Consider, for example, a protein with $N$ amino acids.  The structure and function of the protein is determined by its sequence, but these often are robust to small changes in sequence.   More profoundly, a large family of proteins may share essential structural and functional features while having widely divergent sequences.  We would like to have a description of this ensemble of sequences, ideally being able to write down the probability distribution out of which functional sequences are drawn.  Recent work argues that an effective description of this sequence ensemble for a protein family can be obtained by taking account of pairwise correlations among amino acids at different sites, but ignoring all higher order effects \cite{socolich+al_05,russ+al_05}.  Although this work does not provide an explicit construction of the underlying distribution, it does provide a Monte Carlo procedure for generating new sequences that are consistent with the pairwise correlations in known families, and sequences generated in this way have been proven experimentally to be fully functional. 

What seems like a very different problem is provided by networks of neurons.  If we look in a small window of time, then each neuron either does or does not generate an action potential (spike).  For two cells chosen at random out of a densely connected collection of neurons, one typically finds that pairwise correlations are weak but statistically significant.  Recently it has been suggested that the full pattern of correlations among all the neurons in such a network can be described by the maximum entropy model \cite{jaynes_57,schneidman+al_03} that is consistent with the observed pairwise correlations, and this approach has been shown to provide successful predictions for the combinatorial patterns of activity in the vertebrate retina as it responds to natural movies \cite{schneidman+al_06,tkacik+al_06}.  These maximum entropy models in fact are Ising models with pairwise interactions, which have long been discussed as schematic models for neural networks \cite{hopfield_82,amit_89}; here the Ising model emerges directly as the least structured model consistent with the experimentally measured patterns of coincident spiking among pairs of cells.

These two different examples represent two very different implementations of the idea that complex structures can emerge from pairwise interactions.  It is important to note that in neither case is there any hope that the pairwise description is microscopically exact.  Thus the apparent success of the pairwise approximation provides a first hint that these systems, despite their complexity, are simpler than they might have been.  This idea has been reinforced by yet more recent application of the pairwise, maximum entropy models to other neural systems \cite{shlens+al_06,cosyne07,other_neurons_2}, to a kinase cascade network \cite{tkacik_07}, and to the patterns of gene expression in yeast \cite{psu_group}.

Biochemical, genetic and neural networks all have different structures, and the `networks'   of amino acids in a protein are yet more different.  The mathematical approaches taken in Refs \cite{lockless+ranganathan_99,socolich+al_05,russ+al_05} and \cite{schneidman+al_06,tkacik+al_06}  also are very different, although the theme of pairwise correlations runs through both analyses.  Here we show that the mathematical differences are only differences of emphasis:  In the relevant limit, the Monte Carlo methods of Refs \cite{lockless+ranganathan_99,socolich+al_05,russ+al_05} generate samples drawn out of the maximum entropy probability distribution that would be constructed using the methods of Refs \cite{schneidman+al_06,tkacik+al_06}.  Thus we have a unified framework for exploring the potential of pairwise interactions to tame the complexity of these and other biological systems.  Within this framework we identify some open problems, and comment on the implicit analogies between neural networks and proteins.

\section{Setting up the problem}

Let the system we are studying be described by variables $\sigma_{\rm i}$ that are associated with each element or site ${\rm i}$, where ${\rm i} = 1, 2, \cdots, N$.  For a network of neurons, $\rm i$ can label the individual cells, and $\sigma_{\rm i}$ marks whether that cell generated an action potential in a small window of time; taken together, the set $\sigma_1, \sigma_2 , \cdots , \sigma_N \equiv \{\sigma_{\rm i}\}$ defines the pattern of spiking and silence across the whole network.  For a protein, ${\rm i}$ is an index into the amino acid sequence, and $\sigma_{\rm i}$ indicates which amino acid is found at site ${\rm i}$ along this sequence; the full sequence is defined by $\{\sigma_{\rm i}\}$ \cite{AAnote}.
 
We will phrase our discussion in terms of  ``operators'' $\hat O_\mu  (\{\sigma_{\rm i}\})$ on the set of variables $\{\sigma_{\rm i}\}$. The simplest operators are the variables $\sigma_{\rm i}$ themselves. For neurons, knowing the expectation values $\langle \sigma_{\rm i}\rangle$ means that we know the probability of cell $\rm i$ generating an action potential in a small window of time---the ``firing rate'' of the cell.  For a protein, knowing $\langle\sigma_{\rm i}\rangle$ means that we know the probability of finding each of the twenty possible amino acids at position $\rm i$ in the sequence.  The next most complicated operators involve pairs of variables; knowing the expectation values of these operators corresponds to knowing the probability of two cells generating synchronous action potentials, or the joint probabilities of finding two amino acids at particular locations along the protein sequences.  The central claim of the recent work reviewed above is that knowledge of  the expectation values $\langle \hat O_\mu\rangle$ for these ``one body'' and ``two body'' operators is sufficient to describe, at least to a good approximation, the functional biological system.

One approach to using knowledge of the expectation values $\langle \hat O_\mu\rangle$ is to construct a  probability distribution for the states of the system,
$P(\{\sigma_{\rm i}\})$ that is consistent with this knowledge but otherwise is as random or unstructured as possible; this is the maximum entropy distribution \cite{jaynes_57}. The form of this distribution is given by 
\begin{equation}
P(\{\sigma_{\rm i}\}) = {1\over{Z(\{g_\mu\})}}\exp\left[ -\sum_{\mu=1}^K g_\mu \hat O_\mu  (\{\sigma_{\rm i}\})\right],
\label{Pmaxent}
\end{equation}
where the partition function $Z$ serves to normalize the distribution;
\begin{equation}
Z(\{g_\mu\}) = \sum_{\{\sigma_{\rm i}\}} \exp\left[ -\sum_{\mu=1}^K g_\mu \hat O_\mu  (\{\sigma_{\rm i}\})\right] .
\label{Z}
\end{equation}
By analogy with statistical mechanics it is useful to define the free energy
\begin{equation}
F(\{g_\mu\}) = -\ln Z (\{g_\mu\}).
\end{equation}
Note that there is no real temperature in this problem, or equivalently we have chosen units in which $k_B T = 1$.
The coupling constants $g_\mu$ have to be chosen so that the expectation values in this distribution are equal to their known values, which is equivalent to solving the equations
\begin{equation}
{{\partial F}\over{\partial g_\mu}} = \langle \hat O_\mu \rangle .
\label{fix_avgs}
\end{equation}
This maximum entropy approach is the one used in recent work on the analysis of correlations in networks of neurons \cite{schneidman+al_06,tkacik+al_06}.

As an alternative to the maximum entropy construction, imagine that we create $M$ copies of the system, with variables $\{\sigma_{\rm i}^{(1)}\}, \{\sigma_{\rm i}^{(2)}\}, \cdots , \{\sigma_{\rm i}^{(M)}\}$.  We can evaluate the empirical expectation values of the operators $\hat O_\mu$ across these $M$ copies,
\begin{equation}
\langle \hat O_\mu \rangle_{\rm emp} = {1\over M}\sum_{n=1}^M \hat O_\mu (\{\sigma_{\rm i}^{(n)}\}) .
\end{equation}
Given each of these empirical expectation values, we can try to form a measure of how close these $M$ systems are to being representative of the true expectation values.  Consider
\begin{equation}
\chi^2 = {1\over 2} \sum_{\mu=1}^K W_\mu [\langle \hat O_\mu \rangle_{\rm emp} - \langle \hat O_\mu \rangle]^2 ,
\label{chisq}
\end{equation}
where the $W_\mu$ are weights, expressing how seriously we take deviations in each of the individual operators.  Notice that $\chi^2$ is  a function of all $M\times N$ variables 
$\{\sigma_{\rm i}^{(1)},  \sigma_{\rm i}^{(2)} , \cdots ,  \sigma_{\rm i}^{(M)}\} $.  We can try to force these variables to be representative of the expectation values $\langle \hat O_\mu \rangle$ by drawing from the probability distribution
\begin{equation}
P(\{\sigma_{\rm i}^{(1)},  \sigma_{\rm i}^{(2)} , \cdots ,  \sigma_{\rm i}^{(M)}\} )
= {1\over{{\cal Z}_M}} \exp\left[ - {1\over T} \chi^2 \right] ,
\label{Pannealing}
\end{equation}
and then letting $T\rightarrow 0$; again, ${\cal Z}_M$ is a partition function that serves to normalize the distribution. This annealing procedure is the one used in recent work on the synthesis of artificial proteins \cite{lockless+ranganathan_99,socolich+al_05,russ+al_05}.

\section{Mathematical equivalence of the two methods}

Our goal is to show that the probability distribution for $M$ copies, Eq (\ref{Pannealing}), really is equivalent to the maximum entropy distribution, Eq (\ref{Pmaxent}), in the limit $T\rightarrow 0$ and $M\rightarrow\infty$.  Interestingly, we will see that in this limit the precise values of the weights $W_\mu$ which enter the definition of $\chi^2$ are irrelevant.

To understand the predictions of the annealing method, we need to calculate the partition function ${\cal Z}_M$,
\begin{equation}
{\cal Z}_M = \sum_{\{\sigma_{\rm i}^{(n)}\}} 
\exp\left[ -{1\over T} \chi^2 \right] .
\end{equation}
It will be useful near the end of our discussion to define another free energy $G = -\ln {\cal Z}_M$.  Note that this free energy depends on the expectation values $\{\langle \hat O_\mu\rangle\}$, whereas the free energy $F$ depends on the coupling constant $\{g_\mu\}$. 
 
To make progress we use the standard approach of introducing auxiliary fields $\phi_\mu$ to unpack the quadratic terms in the exponential:
\begin{widetext}
\begin{eqnarray}
\exp\left[ -{1\over T} \chi^2 \right] &=& 
\exp\left[ - {1\over {2T}} \sum_{\mu=1}^K W_\mu [\langle \hat O_\mu \rangle_{\rm emp} - \langle \hat O_\mu \rangle]^2 \right]\\
&=&
\left[\prod_{\mu=1}^K \left(
{T\over{2\pi W_\mu}}\right)^{1/2}\right]
\int d\phi_1 \int d\phi_2  \cdots\int d\phi_ K
\exp\left[
-\sum_{\mu=1}^K {{T\phi_\mu^2}\over{2W_\mu}}
+ i \sum_{\mu=1}^K \phi_\mu \langle \hat O_\mu \rangle_{\rm emp}
- i \sum_{\mu=1}^K \phi_\mu \langle \hat O_\mu \rangle
\right]
\nonumber\\
&&\\
&=& 
\left[\prod_{\mu=1}^K \left(
{T\over{2\pi W_\mu}}\right)^{1/2}\right]
\int d^K\phi 
\exp\left[
-\sum_{\mu=1}^K {{T\phi_\mu^2}\over{2W_\mu}}
- i \sum_{\mu=1}^K \phi_\mu \langle \hat O_\mu \rangle
\right]
\exp\left[ + {i\over M} \sum_{n=1}^M\sum_{\mu=1}^K \phi_\mu  \hat O_\mu (\{\sigma_{\rm i}^{(n)}\})
\right] .
\nonumber\\
&&
\end{eqnarray}
Note that only the last term under the integral depends on the variables $\{\sigma_{\rm i}^{(n)}\}$.
Thus when we compute the partition function we can take the sum over these variables under the integral and write
\begin{eqnarray}
{\cal Z}_M &=& \sum_{\{\sigma_{\rm i}^{(n)}\}} 
\exp\left[ -{1\over T} \chi^2 \right] \nonumber\\
&=&
\left[\prod_{\mu=1}^K \left(
{T\over{2\pi W_\mu}}\right)^{1/2}\right]
\int d^K\phi
\exp\left[
-\sum_{\mu=1}^K {{T\phi_\mu^2}\over{2W_\mu}}
- i \sum_{\mu=1}^K \phi_\mu \langle \hat O_\mu \rangle
\right]
\sum_{\{\sigma_{\rm i}^{(n)}\}} 
\exp\left[ + {i\over M} \sum_{n=1}^M\sum_{\mu=1}^K \phi_\mu  \hat O_\mu (\{\sigma_{\rm i}^{(n)}\})
\right] .
\nonumber\\
&&
\end{eqnarray}
The crucial piece of this equation is the sum over all possible states of the $M$ copies of the system, but since the $M$ copies are independent given $\{\phi_\mu\}$, we can simplify:
\begin{equation}
\sum_{\{\sigma_{\rm i}^{(n)}\}} 
\exp\left[ + {i\over M} \sum_{n=1}^M\sum_{\mu=1}^K \phi_\mu  \hat O_\mu (\{\sigma_{\rm i}^{(n)}\})
\right]
=
\left(\sum_{\{\sigma_{\rm i}\}} 
\exp\left[ + {i\over M} \sum_{\mu=1}^K \phi_\mu  \hat O_\mu (\{\sigma_{\rm i}\})
\right]\right)^M .
\end{equation}
Now we notice that the sum over states in this expression is just the partition function of the maximum entropy distribution, Eq (\ref{Z}), if we identify $-i\phi_\mu/M = g_\mu$.
In this way we can relate the partition function for the annealing problem to an integral over the partition function of the maximum entropy problem,
\begin{equation}
{\cal Z}_M =
\left[\prod_{\mu=1}^K \left(
{T\over{2\pi W_\mu}}\right)^{1/2}\right]
\int d^K\phi 
\exp\left[
-\sum_{\mu=1}^K {{T\phi_\mu^2}\over{2W_\mu}}
- i \sum_{\mu=1}^K \phi_\mu \langle \hat O_\mu \rangle
\right]
\left[ Z(\{ g_\mu = -i\phi_\mu/M\}) \right]^M .
\label{int3}
\end{equation}

The form of Eq (\ref{int3}) suggests that we change variables from $\phi_\mu$ to the coupling constants $g_\mu$, and, recalling that  $Z(\{g_\mu\}) = \exp[-F(\{g_\mu\})]$, we obtain
\begin{eqnarray}
{\cal Z}_M &=&
\left[\prod_{\mu=1}^K \left(
{T\over{2\pi W_\mu}}\right)^{1/2}\right]
(iM)^K
\int d^K g
\exp\left[
+\sum_{\mu=1}^K {{M^2T g_\mu^2}\over{2W_\mu}}
+M  \sum_{\mu=1}^K g_\mu \langle \hat O_\mu \rangle
-M F(\{g_\mu\})
\right]
\\
&=&
\left[\prod_{\mu=1}^K \left(
{{\tilde T}\over{2\pi MW_\mu}}\right)^{1/2}\right]
(iM)^K
\int d^K g
\exp\left[ -M {\cal F}(\{g_\mu\}; \{W_\mu\}; {\tilde T})\right] ,
\label{int5}
\end{eqnarray}
\end{widetext}
where the effective free energy
\begin{equation}
{\cal F}  = 
F(\{g_\mu\}) - \sum_{\mu=1}^K g_\mu \langle \hat O_\mu \rangle
- \sum_{\mu=1}^K {{{\tilde T} g_\mu^2}\over{2W_\mu}} ,
\end{equation}
and ${\tilde T} = MT$.
The partition function of the annealing problem involves an integral over coupling constants, and the integrand is the exponential of $M$ times a free energy that is of order unity as the number of copies $M$ becomes large.  Thus, as $M\rightarrow\infty$, the integral should be dominated by the saddle point where $\partial {\cal F}/\partial g_\mu = 0$ for all $\mu$, or equivalently by values of the coupling constants such that
\begin{equation}
{{\partial F}\over{\partial g_\mu}} = \langle\hat O_\mu \rangle + {{\tilde T}\over{W_\mu}}g_\mu .
\end{equation}
If we consider the limit ${\tilde T}\rightarrow 0$, then this equation is exactly the same as Eq (\ref{fix_avgs}) which sets the values of the coupling constants in the maximum entropy approach.  Thus we can write the saddle point approximation to ${\cal Z}_M$ as
\begin{equation}
{\cal Z}_M \approx A \exp\left[ -M F(\{g_\mu^*\} ) +M \sum_{\mu=1}^K g_\mu^* \langle O_\mu\rangle \right] ,
\end{equation}
where $\{g_\mu^*\}$ are the coupling constants which provide the solution to Eq (\ref{fix_avgs}), and $A$ is a constant that does not depend on $M$.
Finally, we can extract the large $M$ behavior of the free energy $G = -\ln {\cal Z}_M$,
\begin{equation}
\lim_{M\rightarrow\infty} {1\over M} G(\{ \langle \hat O_\mu\rangle\})
=  F(\{g_\mu^*\} ) - \sum_{\mu=1}^K g_\mu^* \langle O_\mu\rangle .
\label{ans}
\end{equation}

To understand the result in Eq (\ref{ans}), we should step back to our original formulation of the two methods.  The free energy $G$ in the annealing method refers to $M$ copies of the system, so it is not surprising that it is proportional to $M$; once we take out this factor we {\em almost} get the free energy $F$ from the maximum entropy method.  The difference is that in the maximum entropy formulation the behavior of the system is a function of the coupling constants $g_\mu$, while in the annealing method the free energy is explicitly a function of the expectation values $\langle\hat O_\mu\rangle$.  This is exactly the situation for the Helmholtz and Gibbs free energies in thermodynamics---the Helmholtz free energy is a function of the volume, and the Gibbs free energy is a function of the pressure.  More generally, whenever we have conjugate variables (pressure and volume, particle number and chemical potential, ... ), we  use the Legendre transformation to connect descriptions based on one or the other member of the conjugate pair \cite{callen_60}.   

For the example of pressure and volume, we have
$G(T,p) = F + pV$ and hence the familiar differential relations
\begin{eqnarray}
{{\partial F(T,V)}\over{\partial V} } &=& -p\\
{{\partial G(T,p)}\over{\partial p} } &=& V.
\end{eqnarray}
Crucially, $F$ and $G$ are descriptions of the same physical system.  More strongly, for large systems (here, $M\rightarrow \infty$) we know that constant pressure and constant volume ensembles are equivalent.
For our problem, the analogous equations are
\begin{eqnarray}
{{\partial F(\{g_\mu\})}\over{\partial g_\mu} } &=& \langle \hat O_\mu\rangle\\
{1\over M}{{\partial G(\{\langle \hat O_\mu\rangle\})}\over{\partial \langle \hat O_\mu\rangle} } &=& -g_\mu.
\end{eqnarray}
The conclusion is that the annealing method, in the $M\rightarrow\infty$, ${\tilde T}\rightarrow 0$ limit, describes $M$ independent copies of the maximum entropy model.

\section{Should we be surprised?}

The derivation above is a bit circuitous, although it does make the connections between the two approaches explicit.  We can make a somewhat shorter, if less constructive, argument.

The probability distribution used in the annealing calculations, Eq (\ref{Pannealing}), is a Boltzmann distribution and hence a maximum entropy distribution.  More precisely, it is the maximum entropy distribution consistent with some average value of $\chi^2$ between the observed and simulated expectation values; as usual $\langle\chi^2\rangle$ is set by the value of $T$.  When we let $T\rightarrow 0$, the expectation value of $\chi^2$ must approach its minimum value, which is zero, unless there is something very odd about the structure of the phase space.  Thus in the $T\rightarrow 0$ limit, the annealing method generates samples from a maximum entropy distribution in which the expectation values computed from the $M$ simulated copies of the system are exactly equal to the observed values.  Finally, if we let $M\rightarrow\infty$, then what we are simulating is an ensemble of samples in which the selected expectation values exactly match their experimental values, but otherwise the distribution of samples has maximum entropy.    That is, we have drawn samples out of the maximum entropy distribution consistent with the observed expectation values.

The subtlety of this argument, which we hope justifies the longer calculation above, concerns the combination of the limits $M\rightarrow\infty$ and $T\rightarrow 0$ and the role of the weights $W_\mu$ in defining $\chi^2$.  The careful calculation shows that we actually need $\tilde T = MT \rightarrow 0$, which is stronger than one might have thought, and that in this limit the weights are irrelevant.

\section{Finite sample size}

Our discussion thus far has assumed that the expectation values $\langle \hat O_\mu (\{\sigma_{\rm i}\}) \rangle$ are known.    In fact we never know these expectation value exactly, since our inferences are based on a finite data set.  For networks of neurons, if we define our variables $\sigma_{\rm i}$ as the presence or absence of a spike from cell $\rm i$ in a small window $\Delta \tau = 10-20\,{\rm ms}$, then an experiment of $\sim 1\,{\rm hr}$ provides more than $10^5$ samples of the state $\{\sigma_{\rm i}\}$, although of course not all these samples are independent \cite{schneidman+al_06,tkacik+al_06}.
With these relatively large sample sizes, it is plausible that we can approximate expectation values, e.g. of the pairwise correlations among neurons $C_{\rm ij} = \langle \sigma_{\rm i}\sigma_{\rm j}\rangle - \langle \sigma_{\rm i}\rangle\langle\sigma_{\rm j}\rangle$, by the corresponding time averages over the experiment, although of course with very large networks the number of pairs can become comparable to the number of samples and we should be careful.  For proteins, in contrast, we have only a few families with $\sim 10^3$ known sequences, and in many cases one must work from fewer than $100$ examples \cite{lockless+ranganathan_99,socolich+al_05,russ+al_05}.  Correspondingly the question of how to treat the issues of statistical significance in the estimation of of $\langle \hat O_\mu\rangle$ has been much more at the center of the discussion of the protein data.

Taken at face value, the construction of $\chi^2$ in Eq (\ref{chisq}) gives our estimates of different operators different weights $W_\mu$.  One plausible choice for these weights (by analogy with usual construction of $\chi^2$) is to set $W_\mu$ equal to the inverse of the variance in our estimates of $\langle \hat O_\mu\rangle$.  This might suggest that the distribution in Eq (\ref{Pannealing}) really does represent the probability of finding the empirical averages $\langle \hat O_\mu \rangle_{\rm emp}$, with $T$ inversely proportional to the number of independent samples $N_{\rm samp}$  in our original database.  But from the maximum entropy distribution we can compute the expected variance in our estimates of the expectation values, and this is
\begin{equation}
\overline {\left[ \delta\langle \hat O_\mu\rangle_{\rm est} \right]^2} 
= {1\over{N_{\rm samp}}} {{\partial^2 F(\{g_\mu\})}\over{\partial g_\mu^2}} .
\end{equation}
Unfortunately, the same arguments can be used to show that errors in the estimates of the different operators in general are not independent, since
\begin{equation}
\overline {\left[ \delta\langle \hat O_\mu\rangle_{\rm est} \delta\langle \hat O_\nu\rangle_{\rm est} \right]} 
= {1\over{N_{\rm samp}}} {{\partial^2 F(\{g_\mu\})}\over{\partial g_\mu\partial g_\nu}} .
\end{equation}
Thus, while the construction of $\chi^2$ provides a convenient heuristic, it can't really represent the likelihood of measuring empirical expectation values given the true expectation values.  Conveniently, in the limit $M\rightarrow \infty, T\rightarrow 0$, all these concerns disappear and even the precise values of the $W_\mu$ are irrelevant.

But if the construction of $\chi^2$ doesn't capture the statistical significance of our estimated expectation values correctly, what should we do instead?    Once we know that we are looking for the maximum entropy distribution consistent with a certain set of expectation values, we know that the {\em form} of the distribution is given by Eq (\ref{Pmaxent}), and our task is to infer the parameters $\{g_\mu\}$.  If we imagine that we have $N_{\rm samp}$ independent samples, with states $\{\sigma_\mu^{(1)} , \sigma_\mu^{(2)} , \cdots , \sigma_\mu^{(N_{\rm samp})}\}$, then the probability of observing these data is given by
\begin{widetext}
\begin{equation}
P(\{\sigma_\mu^{(1)} , \sigma_\mu^{(2)} , \cdots , \sigma_\mu^{(N_{\rm samp})}\})
= \left[ {1\over{Z(\{g_\mu\})}}\right]^{N_{\rm samp}}
\exp\left[ -\sum_{\mu=1}^K g_\mu \sum_{n=1}^{N_{\rm samp}} \hat O_\mu  (\{\sigma_{\rm i}^{(n)}\})\right] .
\end{equation}
Now if we try to find the parameters $\{g_\mu\}$ by maximizing this probability (maximum likelihood estimation),  we find
\begin{eqnarray}
0 = 
{{\partial \ln P(\{\sigma_\mu^{(1)} , \sigma_\mu^{(2)} , \cdots , \sigma_\mu^{(N_{\rm samp})}\})}\over{\partial g_\mu}}
&=& -N_{\rm samp} {{\partial \ln Z(\{g_\mu\})}\over{\partial g_\mu}}
-  \sum_{n=1}^{N_{\rm samp}} \hat O_\mu  (\{\sigma_{\rm i}^{(n)}\})\\
&=&  N_{\rm samp} {{\partial F(\{g_\mu\})}\over{\partial g_\mu}}
-  \sum_{n=1}^{N_{\rm samp}} \hat O_\mu  (\{\sigma_{\rm i}^{(n)}\})\\
{1\over{N_{\rm samp}}} \sum_{n=1}^{N_{\rm samp}} \hat O_\mu  (\{\sigma_{\rm i}^{(n)}\})
&=& {{\partial F(\{g_\mu\})}\over{\partial g_\mu}} .
\end{eqnarray}
\end{widetext}
We see that this is the same as Eq (\ref{fix_avgs}) if we identify the ``known'' expectation values with the averages over our finite set of samples.  Thus, the maximum entropy construction can be viewed as maximum likelihood inference within a specified class of models, and in this framework many questions about the consequences of finite sample size can be seen as part of the more general problem of learning probabilistic models from data \cite{Hinton,note_samps}.

In  a Bayesian framework, we can construct a probability distribution for the parameters $\{g_\mu\}$ given the data $\{\sigma_\mu^{(n)}\}$.  Exploring this distribution, e.g. by Monte Carlo in parameter space \cite{kinney+al_07}, allows us to assign rigorous errors to our parameter estimates.  More importantly, within a Bayesian framework we can integrate over parameters to determine the likelihood that a given {\em class} of models generates the data  \cite{vijay,mackay_03}; this allows us to compare models in which all interactions are possible with those in which some interactions have been set exactly to zero.
In the context of protein structure, if most interactions can be set to zero then we can envision the crucial amino acids as forming limited networks rather than being distributed throughout the protein \cite{lockless+ranganathan_99,suel+al_03}.  In biochemical, genetic and neural networks, setting many interactions to zero would mean describing these networks by a sparsely interconnected graph.    It should be emphasized that the absence of statistically significant correlations between variables $\sigma_{\rm i}$ and $\sigma_{\rm j}$ does not mean that there is no interaction between these variables.  Although this program has not been carried out in any of the systems studied thus far, the Bayesian approach to model selection should provide a rigorous method for deciding on the number of significant interactions \cite{note_bayes}.

\section{Discussion}

As we collect more and more quantitative data on biological systems, it becomes increasingly urgent to find a theoretical framework within which these data can be understood. In many cases, one approach to this problem involves writing down a probability distribution that describes some network of interacting variables:
\begin{itemize}
\item In the context of protein evolution, we would like to write down the probability that any particular amino acid sequence will arise as a functional protein in a certain family.
\item In the context of neural networks, we would like to write down the probability that the network will exhibit any particular pattern of spiking and silence.
\item In the context of genetic networks, we would like to write down the probability that a cell will exhibit any particular combination of gene expression levels, either under a fixed set of conditions or averages over its lifetime.
\end{itemize}

One might object that such probabilistic descriptions are not consistent with the search for a more `rule based' understanding; surely, for example, some sequences form functional proteins and some do not.  Without entering into a philosophical discussion about whether degrees of functionality can be mapped into probabilities, we note that as the systems we are studying become large, our intuition from statistical mechanics is that the difference between a  description in which states are assigned probabilities (the canonical ensemble) and one in which some states are allowed and the rest are not (the microcanonical ensemble) becomes vanishingly small.    Thus, even if we start with a probabilistic description, once we think about proteins with many amino acids or networks constructed from many neurons or genes, we expect that our descriptions will converge to one in which there is a sharp distinction between allowed and disallowed combinations of the underlying variables.

The fact that we are interested in networks with many variables makes the task of constructing a probabilistic description quite daunting.  With $N$ elements, there are $\sim \exp(\alpha N)$ possible combinations of the underlying variables, where $\alpha$ typically is of order unity.  Obviously no experiment will exhaustively explore this configuration space, just as no experiment exhaustively explores the configuration space of the spins in even a small magnetic grain.  For the magnetic grain, however, there is a limited set of natural macroscopic variables to measure---such as the magnetization, specific heat, and susceptibility---that seem to provide a good characterization of the states available to the system.  Can we hope for some analogous simplification in the context of biological networks?

Quantities such as the magnetization and susceptibility can be written as averages over the Boltzmann distribution.  More precisely, the magnetization describes the average behavior of individual spins, and the susceptibility describes the average behavior of pairs of spins (two--point correlations).    The analogous idea, then, is that our description of biological systems might be simplified by focusing on correlations between pairs of elements, rather than allowing for arbitrarily complex combinatorial interactions among many elements.  This is precisely the strategy adopted in recent work on protein sequences \cite{lockless+ranganathan_99,socolich+al_05,russ+al_05} and on networks of real neurons \cite{schneidman+al_06,tkacik+al_06}.  In each case, although the focus on pairwise interactions was implemented differently, the approach was surprisingly successful, accounting for data far beyond the measured correlations.  What we have shown here is that the approaches which arose in different contexts really are the same, so we have a single strategy for simplifying our description of biological systems that seems to be working at very different levels of organization, from single molecules \cite{lockless+ranganathan_99,socolich+al_05,russ+al_05} to biochemical and genetic networks \cite{tkacik_07,psu_group} up to small chunks of the brain \cite{schneidman+al_06,tkacik+al_06,shlens+al_06,cosyne07,other_neurons_2}.  While much remains to be done to test the limits of this approach, this is a very exciting development.

The annealing approach which was used in the analysis of protein sequences allows us to generate directly new samples from the simplified probabilistic model, without actually constructing the model explicitly.  For proteins, these samples are new molecules that can be synthesized, and this has been the path to experimental test of the focus on pairwise correlations.  One could imagine using the same method for neurons to generate new patterns of spiking and silence in the network, and one could then check that the higher--order correlations in these patterns (beyond the pairwise correlations which are matched exactly) agree with experiment.  

The explicit construction of the maximum entropy model, as has been done in the analysis of neurons, allows us to explore the ``thermodynamics'' of the system.  Questions one can address include whether the space of configurations breaks into multiple basins, and whether the parameters of the biological system are in any sense special, e.g. because they are near a critical point.  Perhaps the most direct question we can address given a maximum entropy model for the distribution of states concerns the entropy itself.  In the context of neurons, this entropy sets the capacity for the system to convey information, whether about the external sensory inputs or about some internal variables such as memories and intentions.  In the context of proteins, this entropy measures the number of possible sequences that are consistent with membership in the particular family of functional proteins that we are studying.  An explicit model for the distribution of sequences within a family  is also the proper tool for assessing the likelihood that a previously uncharacterized protein belongs to this family, a practical problem of central importance in analyzing the growing body of sequence data.

\acknowledgments{We thank E Schneidman, GJ Stephens, and G Tka\v{c}ik for helpful discussions.  WB is grateful to C \& R Miller for providing an excuse to visit Texas.  Work in Princeton was supported in part by NIH Grant P50 GM071508 
and by NSF Grants IIS--0613435 and PHY--0650617.
Work in Dallas was supported in part by the HHMI.}


\begin{thebibliography}{99}
\bibitem{AAnote}
To expand the notation a bit, we could replace $\sigma_{\rm i} \rightarrow \sigma_{\rm i}^x$, where $x=1, 2, \cdots , 20$ provides an index for the type of amino acid at site $\rm i$; $\sigma_{\rm i}^x = 1$ means that site $\rm  i$ has amino acid $x$.  In this formulation, of course, we must separately enforce that only one of the twenty $\sigma_{\rm i}^x$ is nonzero.
\bibitem{note_samps}
Interestingly, maximum likelihood inference treats the empirical averages as if they were the exact correlations.
\bibitem{note_bayes}
The analysis of the kinase network in Ref \cite{tkacik_07} comes close to this full Bayesian analysis.  For this problem, the coupling constants clearly form a multimodal distribution, with a substantial fraction of the possible couplings clustered near zero.  Setting these weak couplings exactly to zero makes almost no difference in the likelihood, so with any reasonable prior we expect that rigorous Bayesian model selection would favor models in which these couplings were absent and the resulting matrix of interactions is sparse.
\end{thebibliography}
\end{document}